\documentclass[journal=macromol,manuscript=article]{achemso}

\usepackage{chemformula} 
\usepackage[T1]{fontenc} 
\usepackage[normalem]{ulem}

\usepackage{subfigure}
\usepackage{graphicx}
\usepackage{rotating} 
\usepackage{color}
\usepackage{lscape}
\usepackage{rotating}
\usepackage{mathtools}
\usepackage{tablefootnote}
\usepackage[para]{threeparttable}
\usepackage{array}
 \usepackage{amsmath,amsthm}
 \usepackage{amssymb}

\author{Ji Woong Yu}
\affiliation{Korea Institute for Advanced Study, Seoul 02455, Korea}
\author{Daeseong Yong}
\affiliation{Korea Institute for Advanced Study, Seoul 02455, Korea}
\author{Bae-Yeun Ha}
\email{byha@uwaterloo.ca}
\affiliation{Department of Physics and Astronomy, University of Waterloo, Waterloo, Ontario N2L 3G1, Canada}
\author{Changbong Hyeon}
\email{hyeoncb@kias.re.kr}
\affiliation[KIAS]
{Korea Institute for Advanced Study, Seoul 02455, Korea}
\date{\today}

\title[Brush-induced depletion interaction]
  {Depletion interaction between cylindrical inclusions in polymer brushes}

\abbreviations{AO}

\begin{document}
\begin{figure}
\centering
\includegraphics[width=10.0cm]{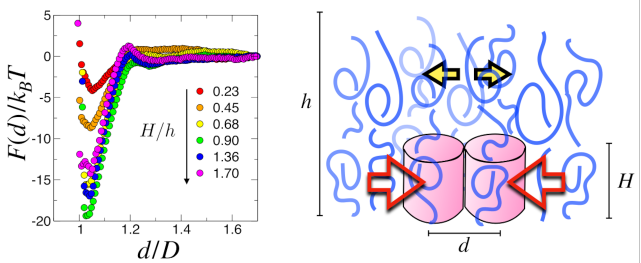}
\captionsetup{labelformat=empty}
\caption{For Table of Contents use only}
\end{figure}
\clearpage 

\setcounter{figure}{0}

\begin{abstract}
Inclusions in mobile brushes experience apparent (depletion) attraction, which arises from the tendency to minimize the volume of depletion zones around the inclusions, thereby maximizing the entropy of the surrounding polymers. 
Here, we study the brush-induced depletion attraction between cylindrical inclusions using molecular dynamics simulations and the Asakura-Oosawa theory.
Our considerations find that the correlation blobs defined in the brush environment serve as the fundamental units of the attraction. 
In tall brushes, however, the entropy of the overgrown polymer competes with the depletion attraction between the inclusions. As a result, 
the brush-induced depletion interaction displays non-monotonic variations with the brush height. 
Our study not only expands the repertoire of colloid-polymer mixtures to depletion interactions in brushes, but also suggests the brush-induced depletion interaction as a previously unappreciated mechanism for glycocalyx-induced protein cluster formation on cell surfaces. 
\end{abstract}
\maketitle

Protein clusters are ubiquitous in cell membranes, and their cellular functions have been a subject of great interest~\cite{aivaliotis2003molecular,lang2010Physiology,baumgart2016NatureMethods,lukevs2017NatComm,sieber2006BJ}. 
For example, an oligomerized form of microbial rhodopsins functions as a light-driven ion-pump or ion-channel in the native membrane~\cite{henderson1975three,klyszejko2008folding,sapra2006characterizing,hussain2015functional,morizumi2019x}. 
It has been shown that the glycocalyx, a layer of glycolipids and glycoproteins that densely coat the cell surface, 
plays vital roles in cell-cell adhesion, communication, and signaling by promoting the integrin nanocluster formation~\cite{offeddu2021cancer,kanyo2020glycocalyx,buffone2019don,shurer2019physical} and regulates the cancer cell progression~\cite{Paszek14Nature}.  
Polymer brush-induced depletion attraction has recently been suggested 
as one of the key driving forces for the transmembrane protein clustering in biological membranes~\cite{tom2021polymer,spencer2021macromolecules}, among other mechanisms, such as membrane undulation-induced thermal Casimir  forces~\cite{goulian1993EPL,park1996JPI,Machta12PRL,spreng2024universal}, protein-protein interaction~\cite{ben1996BJ,schmidt2008PRL,west2009BJ,milovanovic2015NatComm}, and membrane curvature-mediated interaction~\cite{reynwar2007nature,mcmahon2005nature}.

\begin{figure}[t]
    \includegraphics[width=0.55\linewidth]{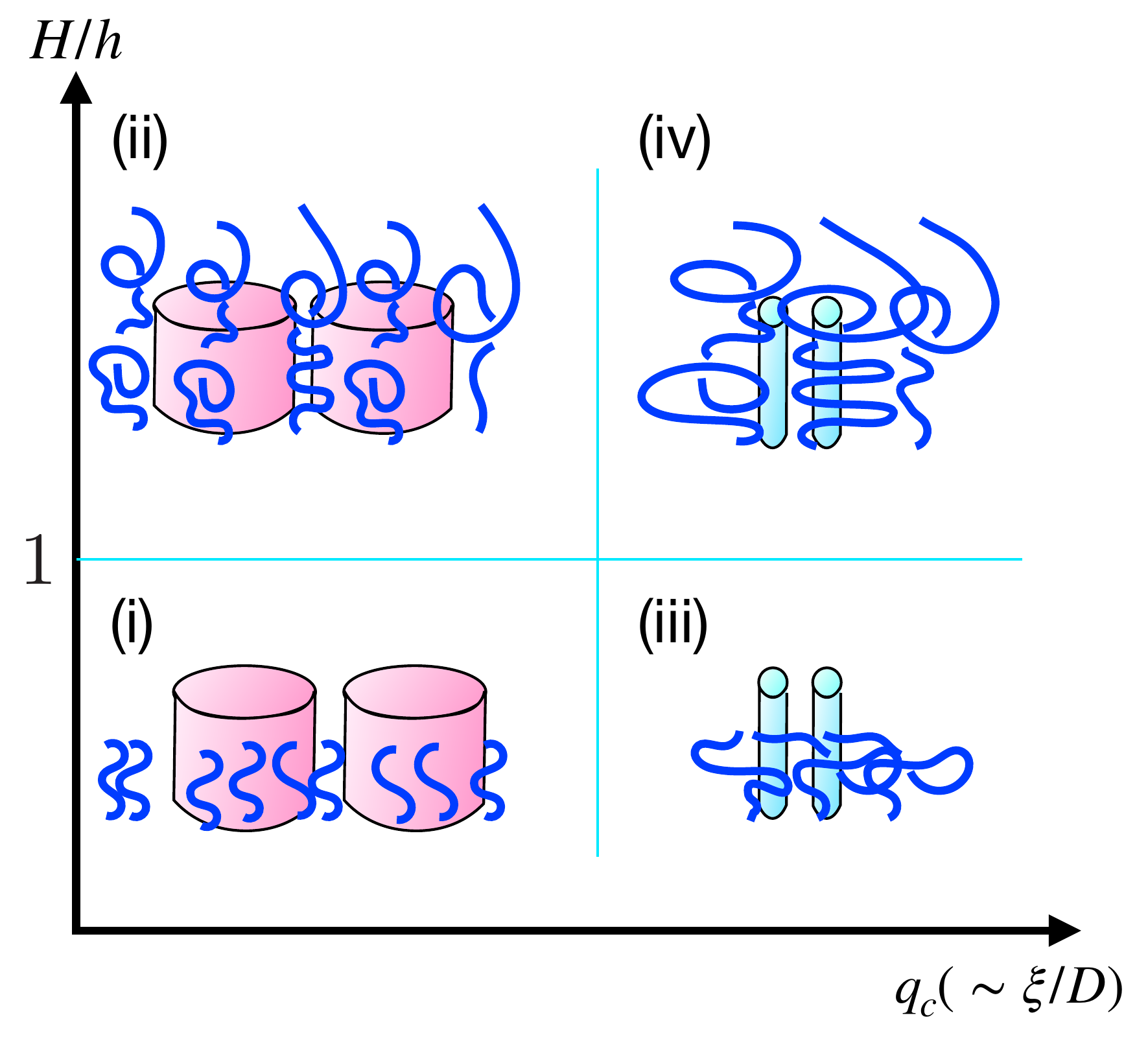}
     \caption{Four distinct regimes of brush-induced depletion interactions between cylindrical inclusions. 
     (i) $H/h<1$, $q_c<1$; (ii) $H/h>1$, $q_c<1$; 
     (iii) $H/h<1$, $q_c>1$; (iv) $H/h>1$, $q_c>1$. 
     }
     \label{fig:4_regimes}
\end{figure}

The attraction between inclusions in a suspension of nonabsorbing depletants characterized with purely repulsive interaction is entropic in nature, arising from an osmotic imbalance of depletants near the depletion layers~\cite{asakura1954JCP,Asakura58JPS}. 
Although the Asakura-Oosawa (AO) theory was originally proposed for interactions between hard spheres or flat surfaces suspended in a solution of depletants in three dimensions, it can straightforwardly be extended to other geometry or polymeric systems as well as to those in two dimensions (2D)~\cite{lekkerkerker2024colloids,miyazaki2022asakura,Kang15PRL,Kang15JACS}. 
For extensively-studied systems of colloid-polymer mixtures~\cite{asakura1954JCP,Asakura58JPS,vrij1976polymers,joanny1979effects,de1981interactions,Shaw91PRA,mao1995depletion,mao1995PRL,biben1996depletion,mao1997JCP,hanke1999polymer,aarts2002phase,Marrenduzzo06JCB,binder2014JCP,lekkerkerker2024colloids,miyazaki2022asakura},
the gyration radius of polymers ($R_g$) and the radius of colloidal particles ($R_c$) serve as the two primary length scales. 
Depending on their size ratio, $q=R_g/R_c$, 
the depletion interactions are either in the \emph{colloid} ($q< 1$)~\cite{eisenriegler1997universal} or in the \emph{protein limit} ($q>1$)~\cite{deGennes1979CRASB,hanke1999polymer,eisenriegler2000polymers}. 
For our cylinder-brush system, on the other hand, where the cylinders can be deemed effectively in a semidilute polymer solution of correlation length $\xi$, $q_c= \xi/D$, 
in which $R_g$ and $R_c$ in $q$ are replaced with
the correlation length $\xi$ and the cylinder diameter $D$, respectively, becomes a relevant parameter. 
Together with $q_c$, due to the presence of another length scale set by 
the brush height ($H$) relative to the inclusion height ($h$), i.e., $H/h$, the brush-induced depletion interactions are formally categorized into four distinct regimes (Fig.~\ref{fig:4_regimes}).   

Here, we first review the basics of brush polymers and revisit a quasi-2D version of the AO theory for the brush-induced depletion interaction~\cite{tom2021polymer}. 
The AO theory is employed to account for the potentials of mean force (PMFs) calculated from molecular dynamics (MD) simulations under varying brush heights and grafting densities. 
We not only highlight the efficacy of the blob concept~\cite{deGennesbook} in quantitative understanding of depletion interactions in a brush environment, but also demonstrate that 
the varying brush height introduces additional complexity to the problem. 
Our study sheds light on membrane biophysics associated with protein nanocluster formation in glycocalyx~\cite{Paszek14Nature,offeddu2021cancer,kanyo2020glycocalyx,buffone2019don,shurer2019physical}. \\

{\bf Polymer brush. }
We consider a polymer brush where $n_p$ polymers, each consisting of $N$ segments, are end-grafted to a 2D surface of area $A$ but laterally mobile on the grafting surface.   
When the grafting density, $\sigma=n_p/A$, is greater than that defined by the Flory radius of an isolated chain ($R_F\simeq bN^{3/5}$), i.e, $\sigma > R_F^{-2}$,  
the polymer chains overlap with each other and transition from a mushroom-like configuration to a string of $N/g$ \emph{correlated} blobs of size each $\xi \, (\simeq \tau^{1/5}bg^{3/5})$, forming a brush of height ($H$) that satisfies the Alexander-de Gennes brush scaling~\cite{Alexander77JP,deGennes1980Macromolecules,de1987polymers}, 
$H\simeq \left(N/g\right)\xi\simeq  Nb(\tau\sigma b^2)^{1/3}$.   
Here, 
$\tau \, (=(T-\Theta)/T\sim\mathcal{O}(1))$ is the relative temperature difference from the theta temperature ($\Theta$) and associated with the second virial coefficient $B_2\sim\tau b^3$, which we set $\tau=1$ for convenience in this paper, 
and $\sigma$ is related to $\xi$ as 
$\sigma\simeq 1/\xi^2$.

The interior of a brush ($\xi<z<H$) is effectively in the semi-dilute regime packed with correlation blobs of size $\xi$ which changes with the monomer volume fraction ($\phi\simeq g/\xi^3$) as   
$\xi \simeq b\phi^{-\gamma}$ with $\gamma=\nu/(3\nu-1)\approx 3/4$ for $\nu=3/5$~\cite{deGennesbook}. 
Thus, the osmotic pressure inside the brush is expected to follow the des Cloizeaux scaling, $\Pi/k_BT\sim 1/\xi^3\sim \phi^{9/4}/b^3\sim (\sigma b^2)^{3/2}$~\cite{deGennesbook,hansen2003osmotic}, {\color{black}where $k_B$ is the Boltzmann constant}. 
\\

{\bf AO theory for brush-induced depletion interaction. } 
For a brush system that contains two parallelly aligned cylindrical inclusions, mimicking signaling transmembrane receptor proteins, e.g., integrins, separated by a center-to-center distance $d$, 
the total volume accessible for brush polymers increases as the inclusions are brought together from $V(d)=V_>$ for $d>D+2\delta_c$ to 
$V(d)=
V_>+V_{\rm ex}(d)$ for $D\leq d\leq D+2\delta_c$ 
with $V_>=(A-\pi D^2/2)\min{(h,H)}\approx A\min{(h,H)}$ and $V_{\rm ex}(d)>0$. 
Then, according to the AO theory, the effective interaction between these inclusions arises from the tendency to minimize the volume of depletion zones, which in turn maximizes the accessible volume for brush polymers to explore. 
Thus, the AO potential is given by 
\begin{align}
\beta F_{\rm AO}(d;\delta_c)&\simeq -n_p \left(\frac{\min{(h,H)}}{\xi}\right)\log{\left[1+\frac{V_{\rm ex}(d;\delta_c)}{V_>}\right]}\nonumber\\
 & \approx -\frac{\sigma}{\xi}V_{\rm ex}(d;\delta_c), 
\label{eqn:dF}
\end{align} 
{\color{black}where $\beta=1/k_BT$.}  
Here, it is conjectured that each correlation blob contributes a free energy of $\sim\mathcal{O}(1)$ $k_BT$ to the depletion interaction. Thus, $n_p\times \min{(h,H)}/\xi$ amounts to the number of blobs in the system {\color{black}characterized by the volume of $A\min{(h,H)}$.} 
The second line of Eq.~\ref{eqn:dF} is obtained from $V_{\rm ex}/V_>\ll 1$ with $\sigma=n_p/A$. 
Our incorporation of the blob idea~\cite{deGennesbook} into the AO theory will be justified through analysis of the numerical results of brush-induced depletion interaction.  

In Eq.~\ref{eqn:dF}, the excess volume for brush polymers 
is equivalent to the excess area $A_{\rm ex}(d;\delta_c)$ multiplied by the height of either brush ($H$) or inclusion ($h$), whichever is smaller, 
i.e.,  $V_{\rm ex}(d;\delta_c)=A_{\rm ex}(d;\delta_c)\min{(h,H)}$ with   
\begin{align}
A_{\rm ex}(x;\lambda_c)=\frac{D^2}{2}
&\left[\left(1+\lambda_c\right)^2\cos^{-1}{\left(\frac{x}{1+\lambda_c}\right)}-x^2\sqrt{\left(\frac{1+\lambda_c}{x}\right)^2-1}\right] 
\label{eqn:Aex}
\end{align}
where we used the rescaled distance $x=d/D$ and thickness $\lambda_c=\delta_c/D$ ($1\leq x\leq 1+\lambda_c$) (see SI for the details of derivation). 
Thus, the AO potential at $x$ reads 
\begin{align}
\beta  F_{\rm AO}(x)
\approx -\sigma \left(\frac{\min{(h,H)}}{\xi}\right) A_{\rm ex}(x;\lambda_c). 
\label{eqn:AO_pot}
\end{align}
In fact, an identical expression can be derived by considering the free energy gain due to the excess volume of $\min{(h,H)}A_{\rm ex}$ created in a  semidilute solution whose osmotic pressure is given by $\beta\Pi\sim 1/\xi^3$,   
\begin{align}
\beta F_{\rm AO}(x)
&\approx -\beta\Pi\times \min{(h,H)} A_{\rm ex}(x;\lambda_c). 
\label{eqn:transfer}
\end{align}

\begin{figure}[t]
    \includegraphics[width=1.0\linewidth]{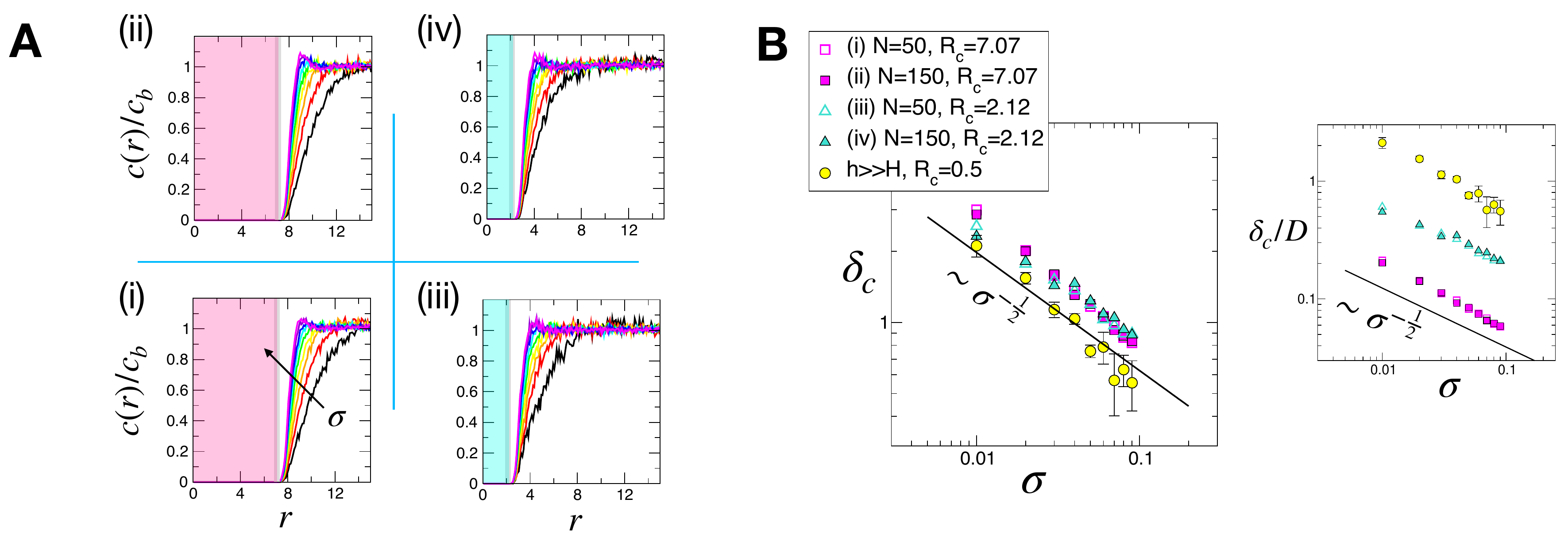}
     \caption{(A) Monomer concentrations around a cylindrical inclusion with $R_c/b=7.07$ (pink box) $R_c/b=2.12$ (pale blue box) obtained from MD simulations in four distinct cases illustrated in Fig.~\ref{fig:4_regimes}. 
     (B) Depletion layer thickness ($\delta_c$) against grafting density ($\sigma$) for the four regimes (i--iv), and those calculated around a needle-like inclusion (yellow circles; a cylinder with $R_c/b=0.5$ and $h\gg H$). 
      The line of $\delta_c\sim\sigma^{-1/2}$ is depicted as a reference. 
     (Inset) Plot of depletion layer thickness rescaled by the cylinder diameter ($\delta_c/D$) against $\sigma$. 
     }
     \label{fig:deltac_regime}
\end{figure}

{\bf Depletion layer thickness. }
To evaluate Eq.~\ref{eqn:AO_pot} or Eq.~\ref{eqn:transfer}, it is necessary to know the depletion layer thickness ($\delta_c$) associated with the parameter $\lambda_c$ in $A_{\rm ex}(x;\lambda_c)$ (Eq.\ref{eqn:Aex}). 
For spheres of radius $R_c$ each in a semidilute polymer solution, it is known that $\delta_c \sim \xi$ {\color{black}for the limit $\xi \ll R_c$ 
~\cite{joanny1979effects,lekkerkerker2024colloids}, whereas for 
the opposite limit $\xi \gg R_c$,} $\delta_c$ is constant as $\delta_c\sim R_c$~\cite{odijk1996protein,sear1997entropy,deGennes1979CRASB}, independent of depletant (polymer) concentration.  
{\color{black}What remains to be clarified is the thickness of a depletion layer around a cylindrical object in a polymer brush or a semidilute polymer solution.  Of particular interest is the case of 
$ R_c<\xi < h$.   At first glance, it appears that $\delta_c$ in this case would combine the features of the two limiting cases for spherical inclusions discussed above.}

For our cylinder-brush system,  we calculate 
$\delta_c$ explicitly from the concentration profiles around a cylinder of radius $R_c \, (=D/2)$ (Fig.~\ref{fig:deltac_regime}) through the  relation~\cite{aarts2002phase} 
\begin{align} 
\pi(R_c+\delta_c)^2=\pi R_c^2+\int_{R_c}^\infty 2\pi r(1-c(r)/c_b)dr,  
\label{eqn:thickness}
\end{align}
where $c(r)/c_b$ is the concentration profile of monomers normalized by the bulk concentration ($c_b\equiv c \, (r >\xi)$) that is reached at a distance of the order of the correlation blob size~\cite{joanny1979effects}. 
For polymers grafted as in brushes, the symmetry along the brush height is broken; however, the brush height-dependence of the concentration profile around the cylinder is found insignificant.
The calculations for $\delta_c$ are carried out over the range of $\sigma b^2 \, (=0.01-0.09)$ and $D/b=1$, $3\sqrt{2}$ and $10\sqrt{2}$, 
so that the parameter $q_c=\xi/D\simeq\delta_c/D\simeq (\sigma^{1/2}D)^{-1}$ encompasses the range of $q_{c,{\rm min}}<q_c<q_{c,{\rm max}}$ with $q_{c,{\rm min}}\simeq 0.235$ for $\sigma b^2=0.09$ and $D/b=10\sqrt{2}$ and $q_{c,{\rm max}}\simeq 10.0$ for $\sigma b^2=0.01$ and $D/b=1$ (Fig.~\ref{fig:deltac_regime}). {\color{black}Throughout the paper, all lengths are measured in units of $b$, unless otherwise stated.}   

First, {\color{black}our explicit calculation using MD simulation results finds that} the depletion layer thickness is narrow compared to the diameter of the cylinder ($\delta_c/D< 1$) over the whole range of $q_c$ being explored ($0.235\leq q_c\leq 10.0$). 
Although the depletion layer thickness around the cylinder is curvature-dependent, such that smaller cylinder curvatures lead to greater $\delta_c$'s,  
the actual difference between $\delta_c$'s for different $D$ is only minor (see Fig.~\ref{fig:deltac_regime}).   
Furthermore, the difference between $\delta_c$'s for the short ($H<h$) and tall brushes ($H>h$) is not statistically significant either (Fig.~\ref{fig:deltac_regime}B). 

{\color{black}Second, it may be tempting to associate  the $q_c>1$ limit of cylindrical inclusions with the small-$R_c$ limit of \emph{spherical} inclusions, where the depletion layer thickness is set by $R_c$, independent of $\xi$~\cite{odijk1996protein,sear1997entropy,deGennes1979CRASB}.}  
However, the scaling relation of $\delta_c\sim \sigma^{-1/2}$ revealed from our calculation (Fig.~\ref{fig:deltac_regime}B) is apparently at odds with such a notion. 
This signifies that the depletion layer thickness around cylindrical inclusion in brushes is dictated by the blob size ($\delta_c\sim\xi$) regardless of the value of $q_c$ (Fig.~\ref{fig:deltac_regime}) and that the correlation blob is 
the fundamental interaction unit for the brush-induced depletion attraction, which should hold for $h>\xi$.  
Note that the cylinder-brush system is distinguished from the colloid-polymer mixtures in that the axial dimension of cylinder ($h$) is still greater than $\xi$ {\color{black}(see below for additional  details)}. 
\\

\begin{figure}[ht!]
    \includegraphics[width=1.0\linewidth]{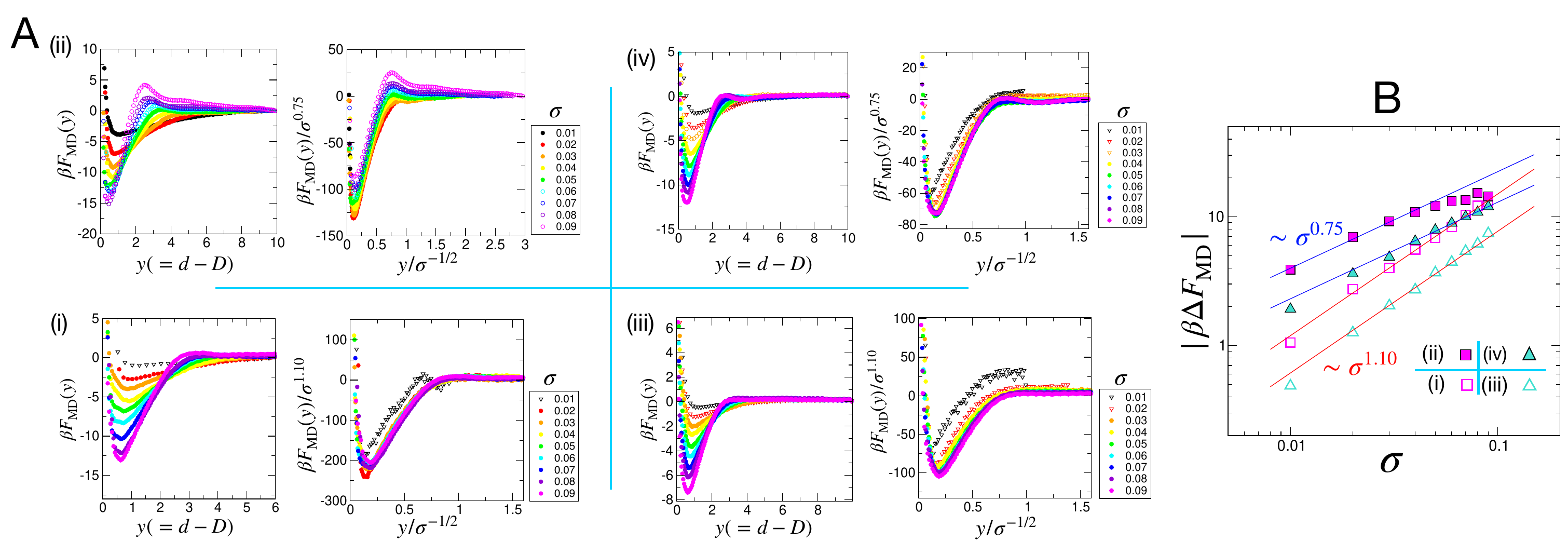}
     \caption{PMFs between cylinders with varying $\sigma$ obtained from MD simulations for the four distinct regimes (i)--(iv). 
     (A) (left) PMFs versus $y=d-D$.  
     (right) PMFs in the range of $\sigma$ represented by the filled circles are collapsed onto one another when the distance and free energy are rescaled, respectively, with $\sigma^{-1/2}(=\xi)$ and $\sigma^{\alpha}$ ($\alpha=1.10$ for $H/h<1$ in (i) and (iii); $\alpha=0.75$ for $H/h>1$ in (ii) and (iv)). 
     (B) $|\beta\Delta F_{\rm MD}|$ versus $\sigma$. 
     The red and blue lines depict the scalings of $\sim\sigma^{1.10}$ and $\sim\sigma^{0.75}$, respectively. 
     }
     \label{fig:collapse}
\end{figure}

{\bf Analysis of PMFs from MD simulations.} 
The PMFs calculated from MD simulations for the four regimes illustrated in Fig.~\ref{fig:4_regimes} are analyzed by using Eq.~\ref{eqn:AO_pot} (see Fig.~\ref{fig:collapse}). 
For $H<h$, $\min{(h,H)}/\xi=H/\xi=N/g$, and Eq.~\ref{eqn:Aex} at $x=1$ for $\lambda_c\ll 1$ is approximated as $A_{\rm ex}(1;\lambda_c)\sim D^2\lambda_c^{3/2}\sim D^{1/2}\delta_c^{3/2}$.     
From the relation of $\delta_c\sim \sigma^{-1/2}$ (Fig.~\ref{fig:deltac_regime}) and the blob concept with $\nu=0.588$, 
the free energy gain, $\beta  F_{\rm AO}(x=1)\equiv \beta\Delta F_{\rm AO}$ (Eq.~\ref{eqn:AO_pot}), is expected to scale with $\sigma$ as 
\begin{align}
|\beta\Delta F_{\rm AO}|
&\sim \sigma\left(\frac{N}{g}\right)D^{1/2}\delta_c^{3/2}\sim 
ND^{1/2}\sigma^{1/4+1/2\nu}\sim \sigma^{1.10}. 
\label{eqn:stability_smallH}
\end{align} 
Fig.~\ref{fig:collapse}B shows that 
the dependence of stability on $\sigma$ predicted by Eq.~\ref{eqn:stability_smallH} 
well accounts for the MD simulation results. 
In addition, upon rescaling the inclusion gap ($y=d-D$) by the blob size ($\xi\sim \sigma^{-1/2}$) and the free energy by the $\sigma$-dependent stability (Eq.~\ref{eqn:stability_smallH}), the rescaled free energy profiles with varying $\sigma$ 
overlap nicely with each other (see Fig.~\ref{fig:collapse}A-(i) and (iii)), except for those with small $\sigma$ (the empty symbols, $\sigma=0.01$ in Fig.~\ref{fig:collapse}A-(i) and $\sigma=0.01$, 0.02 in Fig.~\ref{fig:collapse}A-(iii)) which are on the border of the brush forming regime ($\sigma R_F^2\gtrsim 1$).  
On the other hand, for $H>h$,  
the free energy gain 
is characterized by a distinct exponent, $3/4$,  as follows: 
\begin{align}
|\beta\Delta F_{\rm AO}|
&\sim \frac{1}{\xi^3}\times h D^{1/2}\delta_c^{3/2}\sim hD^{1/2}\xi^{-3/2}\sim \sigma^{3/4}. 
\label{eqn:stability_largeH}
\end{align} 

\begin{figure}[ht!]
    \includegraphics[width=0.6\linewidth]{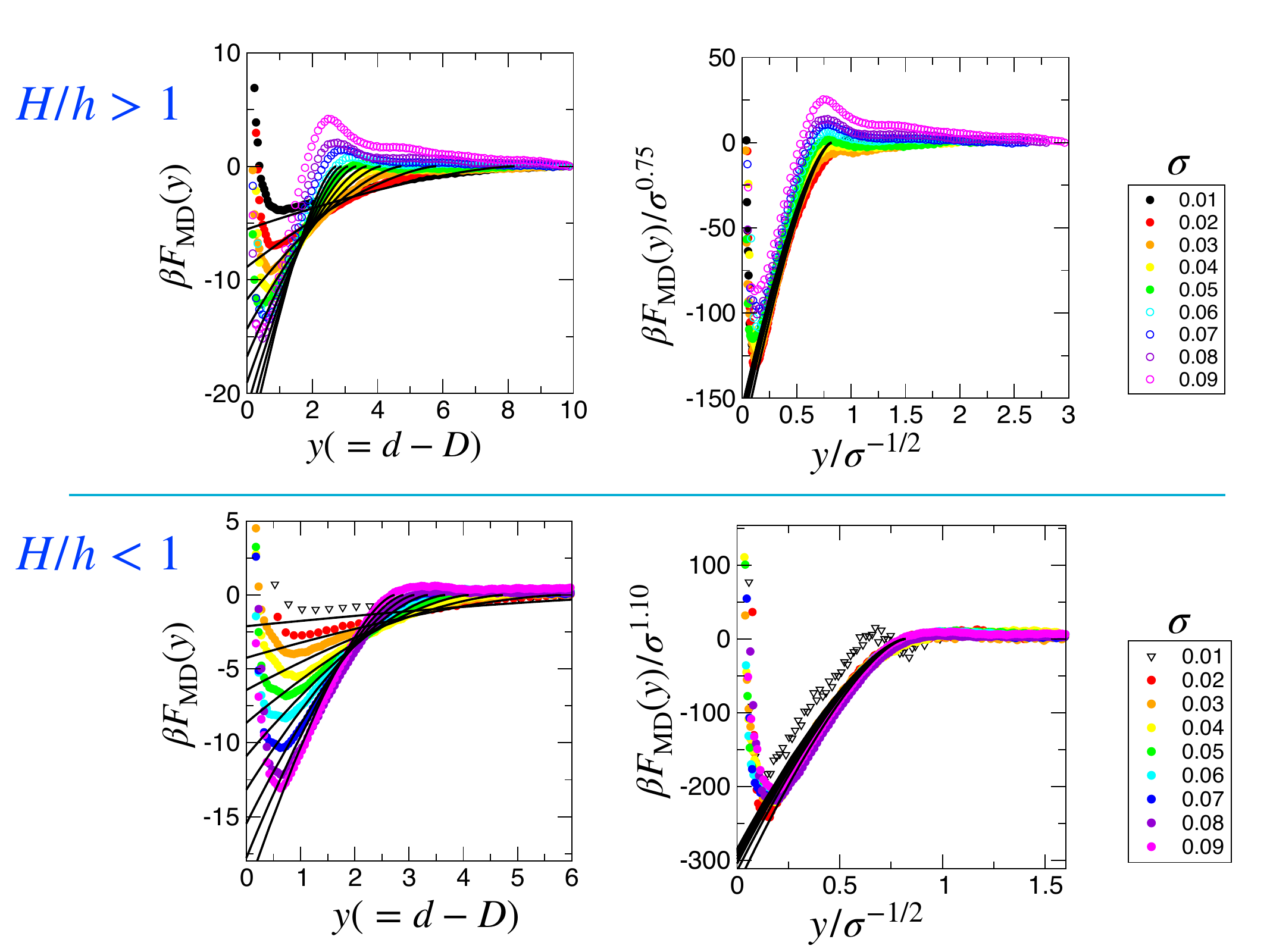}
     \caption{AO potentials, $\beta F_{\rm AO}(y)$ (Eq.~\ref{eqn:AO_pot} with Eq.~\ref{eqn:Aex}, black lines) with $D/b=10\sqrt{2}$ and $h/b=20\sqrt{2}$ for $N=150$ ($H/h>1$, top) and $N=50$ ($H/h<1$, bottom) after rescaling $y\rightarrow \tilde{y}\simeq y/1.22$ and $\beta F_{\rm AO}(y)\rightarrow \beta \tilde{F}_{\rm AO}(\tilde{y})\simeq 1.43\times \beta F_{\rm AO}(\tilde{y})$ over all $\sigma$'s{\color{black}, where the numerical prefactors $1/1.22$ and $1.43$ stem from the scaling relation without prefactor ($\xi \sim \sigma^{-1/2}$) used in deriving the $\sigma$-dependent scaling of AO potentials.}
     The $\beta \tilde{F}_{\rm AO}(\tilde{y})$ for varying $\sigma$'s are overlaid on top of PMFs obtained from MD simulations ($\beta F_{\rm MD}(y)$). 
     The panels on the right show the collapsed AO potentials after rescaling $\tilde{y}\rightarrow \tilde{y}/\sigma^{-1/2}$ and $\beta \tilde{F}_{\rm AO}(\tilde{y})\rightarrow \beta \tilde{F}_{\rm AO}(\tilde{y})/\sigma^\alpha$ with $\alpha=1.10$ ($H/h<1$) and $\alpha=3/4$ ($H/h>1$). 
     }
     \label{fig:AO_MD_direct}
\end{figure}

{\color{black}Upon rescaling the inter-cylinder gap ($d-D$) by the $\sigma$-dependent blob size, and the free energy by the $\sigma$-dependent stability of the AO potential, i.e.,  
\begin{align}
(d-D)&\rightarrow (d-D)/\xi\sim (d-D)/\sigma^{-1/2}\nonumber\\
\beta F_{\rm MD}(y)&\rightarrow \beta F_{\rm MD}(y)/|\beta \Delta F_{\rm AO}|\sim\beta F_{\rm MD}(y)/\sigma^{\alpha} 
\end{align} 
where $\alpha=1.10$ for $H<h$ and $\alpha=3/4$ for $H>h$, 
the free energy profiles generated at different $\sigma$ collapse onto each other (see Fig.~\ref{fig:collapse}A). 
Furthermore, Fig.~\ref{fig:AO_MD_direct} shows that  
the AO potentials (Eq.~\ref{eqn:AO_pot}) calculated using the explicit expression of $A_{\rm ex}(x)$ (Eq.~\ref{eqn:Aex}) are nicely overlaid, at least, onto the lower part of the PMFs from MD simulations, i.e., $\beta F_{\rm MD}(y)<0$. 
The repulsive barriers that appear in $\beta F_{\rm MD}(y)>0$ at large $\sigma$'s in  Fig.~\ref{eqn:AO_pot}A are the most pronounced at short inter-cylinder separation ($y\approx (2-3)$ or $y/\sigma^{-1/2}\approx 0.75$) for the case of thick cylinders in tall brushes (compare the case (ii) with others in Fig.~\ref{fig:collapse}A). 
The repulsive barriers in $\beta F_{\rm MD}(y)$'s can be attributed to the formation of depletion zones above the inclusion, which will be discussed further in the section that follows. 
The collapse of PMFs upon the rescaling in Fig.~\ref{fig:collapse}A, the $\sigma$-dependent scaling of the free energy gain quantitatively confirmed in Fig.~\ref{fig:collapse}B, and the quantitative agreement between the AO potential and PMFs from MD for varying $\sigma$'s highlighted in Fig.~\ref{fig:AO_MD_direct} lends support to our analysis based on the AO theory. 
} 

\begin{figure}[ht!]
    \includegraphics[width=0.7\linewidth]{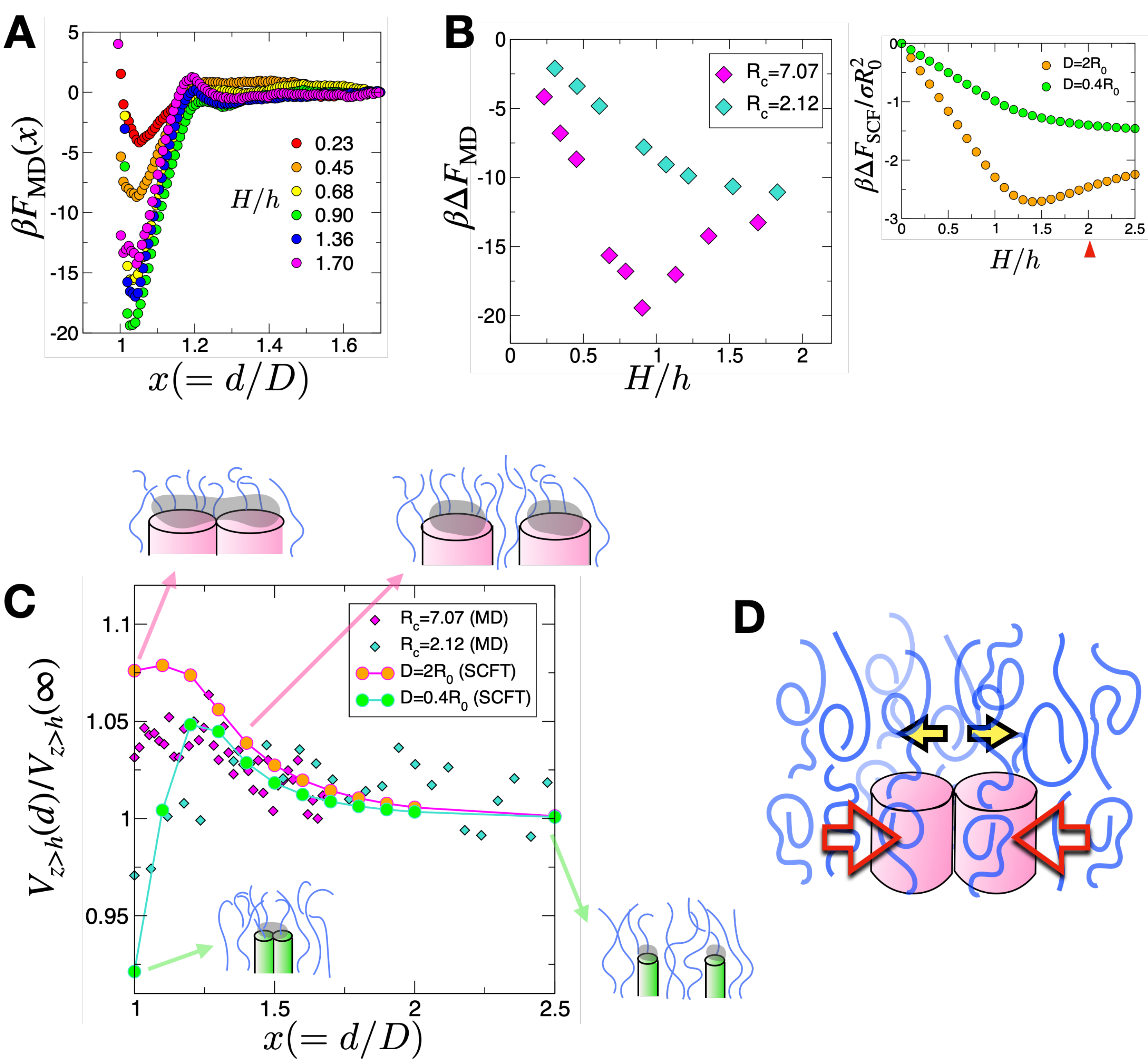}
     \caption{(A) PMFs from MD simulations for $R_c/b=7.07$ at $\sigma b^2=0.08$ with varying $H/h$. 
     (B) Stabilities of two cylinders as a function of $H/h$ for $R_c=7.07$ and $2.12$ at $\sigma b^2=0.08$. 
     (Inset) Self-consistent field (SCF) calculation{\color{black}s for the diameters, $D=2R_0$ and $0.4R_0$,} with the excluded volume parameter $\Lambda=2\pi^2$ (see SI). 
     (C) The depletion zone volume above the cylinders with increasing distance between the cylinders, $V_{z>h}(d)/V_{z>h}(\infty)$, obtained from both MD simulations (diamonds) and SCF calculations (circles) at $H/h=2$.  
     The volume, $V_{z>h}(d)$, is obtained by integrating the $z$-dependent volume fraction of monomers at the bulk ($\phi_b(z)$) subtracted by that at ${\bf r}$ over the space ($\phi({\bf r};d)$), i.e., $V_{z>h}(d)=\int [\phi_b(z)-\phi({\bf r};d)]d{\bf r}$.
     The cartoons depict the depletion zone volume (gray) above the {\color{black}thick and thin} cylinders. 
     (D) An illustration of depletion attraction (red arrows) and repulsion (yellow arrows) between two {\color{black}thick cylinders} in tall brushes.}
     \label{fig:PMF}
\end{figure}

The depletion potentials and stabilities for brushes spanning from $H<h$ to $H>h$ are calculated for a fixed $\sigma$ as well (Fig.~\ref{fig:PMF}).   
As predicted by the AO theory 
(Eq.~\ref{eqn:stability_smallH}), 
a cluster of inclusions is further stabilized with an increasing brush height $H$ (or $N$) as long as $H<h$. 
For $H>h$, however, reduction of the stability is significant ($H/h\gtrsim 1.5$), 
especially for the inclusions with large $R_c \, (=7.07)$ (Fig.~\ref{fig:PMF}B), which is no longer explicit in Eq.~\ref{eqn:stability_largeH} although it  correctly captures the $\sigma$-scaling. 

Calculations of stability between cylinders in brushes based on the self-consistent field approach, $\beta\Delta F_{\rm SCF}/\sigma R_0^2$, where $R_0^2=Nb^2$ (see SI), confirms a similar trend of non-monotonic variation (the inset of Fig.~\ref{fig:PMF}B), validating its thermodynamic origin ruling out a kinetic effect. 
Our inspection of the results from both SCF calculation and MD simulations focusing on the depletion zone (the volume inaccessible to the polymer segments) \emph{above} the cylinders ($V_{z>h}$, depicted in gray in the cartoons of Fig.~\ref{fig:PMF}C)
indicates that the reduction of the depletion zone volume is significant when two cylinders are 
separated apart (Fig.~\ref{fig:PMF}C). 
Just like the depletion attraction between cylinders
that results from the tendency to minimize the depletion zone around the 
cylinder body, repulsion arises while minimizing the excess depletion zone above the cylinders ($z>h$) surrounded by the overgrown polymer segments (the yellow arrows, Fig.~\ref{fig:PMF}D), and it partially offsets the depletion attraction (the red arrows, Fig.~\ref{fig:PMF}D) acting on the main body of the cylinders ($z<h$). 

\begin{figure}[ht!]
    \includegraphics[width=1.0\linewidth]{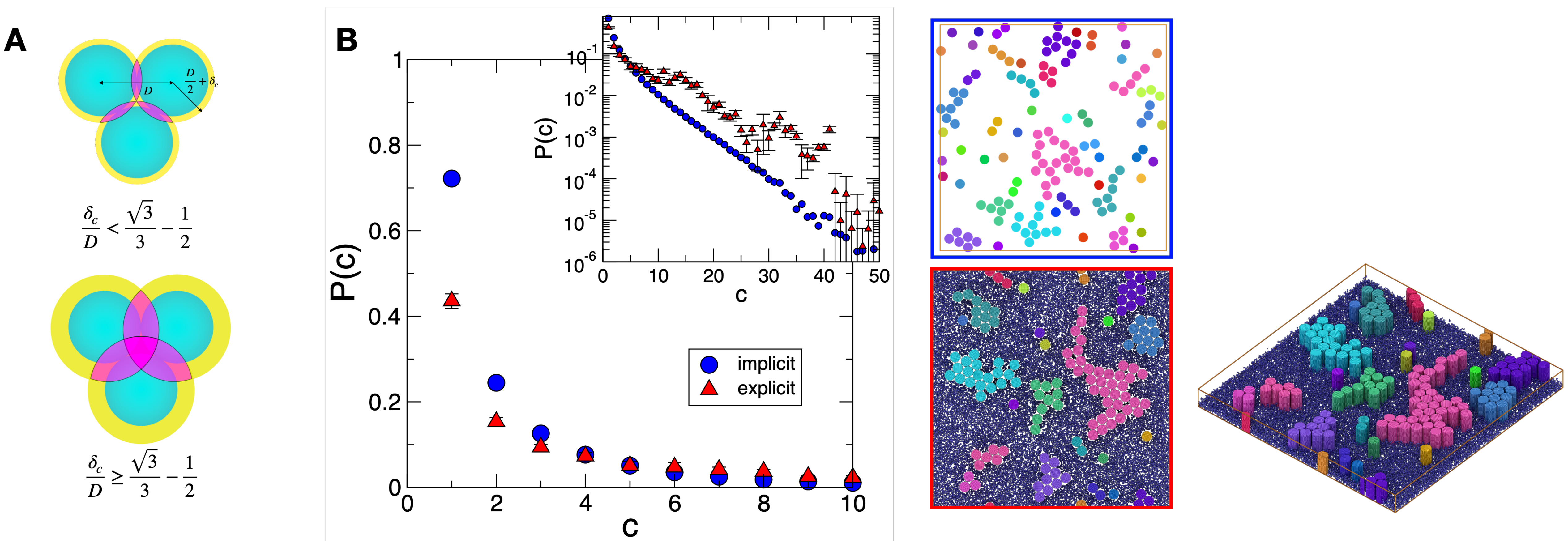}
     \caption{\textcolor{black}{Cluster size distribution, $P(c)$, (i) based on MD PMF between two cylinders (implicit) and (ii) obtained directly from MD simulations (explicit) for cylinders with an area fraction $\phi_c=0.2$ in brushes ($N=20$) at $\sigma b^2=0.06$. For cylinders, we have chosen $D=7.07 \times 2 b$ and $h\simeq28b$. 
     (A) Depletion zone shared between three cylindrical inclusions under the condition of compact packing (magenta) for $\delta_c/D<\sqrt{3}/3-1/2$ (top) and $\delta_c/D\geq\sqrt{3}/3-1/2$ (bottom).  
     (B) The plot of cluster size distribution $P(c)$ and the snapshots of clusters obtained from the implicit (top) and explicit simulations of cylinders in brushes (bottom), clarifying their similarities and differences.} }
     \label{fig:cluster}
\end{figure}

\textcolor{black}{The AO theory incorporating the blob concept and depletion zone volume offers a conceptual framework for interpreting our simulation data.  To further clarify the applicability of the AO theory, we have examined clustering of cylindrical inclusions driven by brush-induced depletion interactions.  It is worth noting that as is most obvious for simple crowders (e.g., hard spheres), 
entropic forces such as depletion forces are generally non-pairwise additive~\cite{lekkerkerker2024colloids,Kang15PRL}. 
Many-body interactions among cylinders and the correlations between brush polymers are not properly taken into account if one were to perform a simulation that employs  
a PMF between two cylinders without explicitly modeling brushes. 
For the brush-induced depletion interaction discussed here, 
it is evident from the geometry that if the ratio between the depletion layer thickness and the cylinder diameter ($\delta_c/D$) exceeds $\approx 8$ \% ($(\delta_c/D)^\ast=\sqrt{3}/3-1/2\approx 0.08$), the volume of the depletion zone shared by three cylinders can no longer be represented as 
a sum of pairwise overlaps between their individual depletion zones (see Fig.~\ref{fig:cluster}A)~\cite{lekkerkerker2024colloids}.}

\textcolor{black}{Figure~\ref{fig:cluster}B shows the distributions of cluster size, $P(c)$, where the cluster size $c$ is defined as the number of cylindrical inclusions in a cluster, obtained from two distinct simulations (see SI for the details):  
(i) An implicit simulation of cylinders (circular disks) with an area fraction $\phi_c=0.2$ without explicit brushes that employs the pairwise PMF obtained through the umbrella sampling in brushes with $N=20$ and $\sigma b^2=0.06$; 
(ii) Brute-force MD simulation of cylinders ($\phi_c=0.2$) inserted to an explicit brush environment ($N=20$) at 
$\sigma b^2=0.06$. 
In both cases, we have chosen $D=7.07 \times 2 b$ and $h\simeq 28b$ for the cylinders.  Also included are the snapshots of clusters obtained from the implicit (top) and explicit simulations of cylinders in brushes (bottom). 
The depletion layer thickness around the cylinders in brushes with $\sigma b^2=0.06$, which can be estimated directly from the inset of Fig.~\ref{fig:deltac_regime}B, is less than the crossover value, $\delta_c/D\approx 0.07 < (\delta_c/D)^\ast$. 
Thus, the implicit and explicit models are expected to be in agreement (Fig.~\ref{fig:cluster}A top). 
At least, for $c\lesssim 10$, the implicit and explicit simulations give rise to comparable cluster size distributions.  For $c>10$, on the other hand,  the explicit simulations tend to produce clusters of greater size  although such a probability is relatively small (see also the inset of Fig.~\ref{fig:cluster}B that plots the $P(c)$ in a logarithmic scale and  compare the snapshots).   
This points to the critical role of many-body correlations, which can be incorporated to improve the current AO framework.  } 

\begin{figure}[ht!]
    \includegraphics[width=0.6\linewidth]{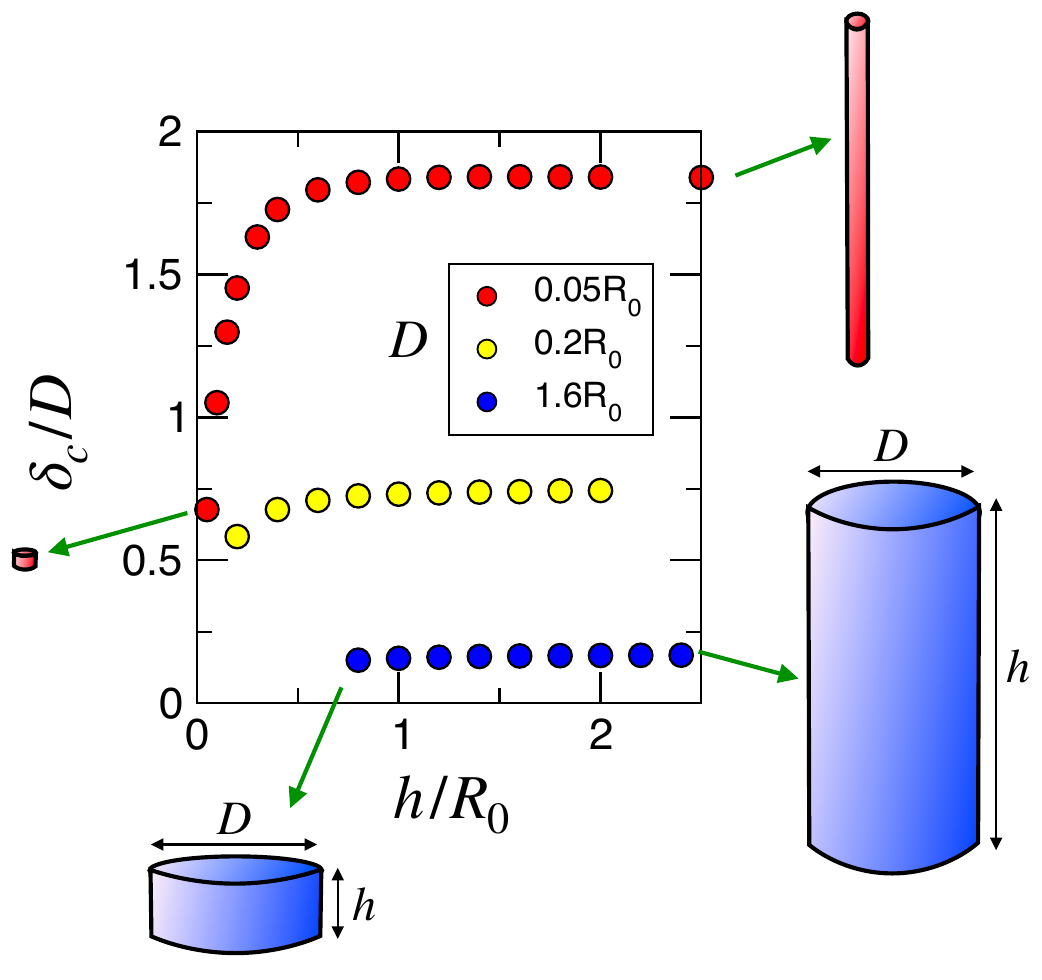}
     \caption{\color{black}Depletion layer thickness ($\delta_c$) around cylindrical inclusions versus their height ($h$). 
     The SCFT calculations with the excluded volume parameter $\Lambda=2\pi^2$ were carried out (see SI) to obtain the density profile around cylinders ($\phi({\bf r})$ in Eq.~S9 in SI, and $\phi({\bf r})=c(r;z)$ due to radial symmetry around cylinders) for different aspect ratios of the cylinder. The value of $\delta_c$ was then calculated using Eq.~\ref{eqn:thickness} at the brush height, $z(<h)$, giving rise to the maximum density profile. 
     For thin cylinders (red circles, $D=0.05 R_0$), the reduced thickness ($\delta_c/D$) initially increases rapidly with $h/R_0$ until it reaches a plateau.
     }
     \label{fig:depletion_h_effect}
\end{figure}
The results from our analysis of brush-induced depletion interactions are recapitulated as follows. 
The AO theory, which relies solely on the entropy argument, i.e., the principle of minimizing the depletion zone around cylindrical objects in brushes, thereby maximizing the volume for brush polymers to explore,  
elucidates the origin of non-monotonic variation of depletion interactions with growing brush height.   
Even when the cylinder is needle-like, the depletion layer thickness is not comparable to the diameter of the cylinder as has been assumed in a previous study on bundle formation of rod-like filaments in a polymer solution~\cite{de2001flexible} based on small-spherical inclusions 
{\color{black}in a semidilute solution}
~\cite{deGennes1979CRASB,odijk1996protein,sear1997entropy}. 
{\color{black}Specifically, de Vries studied the ``insertion free energy'' of a thin cylinder in a polymer solution by considering the cylinder as a linear succession of small spheres of size $R_c<\xi$ and assuming that the depletion layer thickness around the cylinder remains identical to $R_c$~\cite{de2001flexible}.  
However, in addition to Fig.~\ref{fig:deltac_regime}B that shows the scalings of $\delta_c\sim\sigma^{-1/2}$ for both thin and thick cylindrical inclusions, explicit calculations of $\delta_c$ around cylinders with various $D$ based on SCFT 
clearly show significant variation in the depletion layer thickness, especially, when $D$ is small (see Fig.~\ref{fig:depletion_h_effect}). 
For $D=0.05R_0$, the $\delta_c/D$ increases steeply from $\delta_c/D\sim 0.5$ with a growing size of cylinder  ($h$) and saturates to $\delta_c/D\lesssim 2$ when $h\gtrsim R_0$.  
This implies that in clear contradiction to the de Vries' earlier  presumption, the depletion layer thickness surrounding an inclusion increases as one connects small spheres into a cylinder.
These findings can be understood based on the following physical picture: a thin cylinder starts to feel the entirety of blobs if its height gets greater than the blob size ($D< \xi<h$).}
The blob concept is still relevant in accounting for the depletion interaction between rod-like objects in polymer solutions. 
As a result, the concept of correlation blobs incorporated into the AO theory as the fundamental unit of depletion interactions still proves effective in quantitative characterization of 
the depletion layer and stability between two inclusions, making the problem of brush-induced depletion interaction distinct from that of colloid-polymer mixtures. 
\\

{\bf Methods.}
The cylinders with $D=3\sqrt{2}b$, $10\sqrt{2}b$ and $h=20\sqrt{2}b$ were modeled using a composite body.  
The beads at the bottom layer ($z=0.0$) were harmonically restrained along the $z$-direction with a force constant $k\simeq10^3\epsilon/b^2$, where $\epsilon$ is the energy scale of WCA potential described below. 
The brush polymers were modeled by employing  
the potential along the chain, $U_{\rm poly}(r_{i,i+1}) = -(k_\textrm{F}r_o^2/2)\ln\left[1-\left(r_{i,i+1}/r_o\right)^2\right]+4\epsilon\left[(b/r_{i,i+1})^{12}-(b/r_{i,i+1})^6+1/4\right]$ with $k_\textrm{F}=30.0\epsilon/b^2$ and $r_o=1.5b$. 
The monomers comprising the brush polymers and cylinders  
repel through the WCA potential $U_\textrm{WCA}(r)=
    4\epsilon\left[(b/r)^{12}-(b/r)^6+1/4\right]$ for $0<r<2^{1/6}b$ and 
    $U_\textrm{WCA}(r)=0$ otherwise. 
    To prevent polymers from penetrating the surface at $z=0$, the WCA potential was imposed at $z=-b$. 
For a given $\sigma$, $n_p$ chains were grafted to the box that has a dimension of $L_x\times L_y$  ($L_x=L_y=\sqrt{n_p/\sigma}=100b$) with the periodic boundary imposed along the $x$ and $y$ directions. 
Along the $z$ direction, the shrink-wrapped boundary condition was used. 
Two distinct sizes of brush polymer, $N=50$ and $150$, were employed to simulate the regimes of (i), (iii) ($H<h$) and (ii), (iv) ($H>h$) in Fig.~\ref{fig:4_regimes}, respectively.

The Large-scale Atomic/Molecular Massively Parallel Simulator (LAMMPS)~\cite{thompson2022lammps} was used for the MD simulations. 
To calculate the PMF between the cylinders, 
the umbrella sampling was performed at the temperature $T=1.0$ $\epsilon/k_B$, with $k_B$ the Boltzmann constant. 
At each sampling point, the system was first relaxed for $10^3\tau_{\rm MD}$ where $\tau_{\rm MD}\approx (m b^2/\epsilon)^{1/2}$ denotes the intrinsic time scale for MD simulations, followed by a production run for $10^5\tau_{\rm MD}$ under a bias potential, $U_\textrm{b}\left(d;d_j\right)=k_\text{b}\left(d-d_j\right)^2/2$, with $k_\textrm{b}=200.0$ $\epsilon/b^2$ and $d_j=D+10b-jb/4$ ($j=1,...,42$). 
The unbiased free energy profile was reconstructed through the weighted histogram analysis method (WHAM)~\cite{KumarJCC1992}.

\begin{acknowledgement}
This study was supported by the KIAS individual grants, AP091501 (JWY) and CG035003 (CH), the National Research Foundation of Korea (NRF) grant, NRF-2022R1C1C2010613 (DY), funded by the Korea government (MSIT), and by 
Natural Sciences and Engineering Research Council of Canada (B-YH). 
We thank the Center for Advanced Computation in KIAS for providing the computing resources.
\end{acknowledgement}

\clearpage 

\section{Supporting Information}
\setcounter{equation}{0}
\setcounter{figure}{0}
\renewcommand{\theequation}{S\arabic{equation}}
\renewcommand{\thefigure}{S\arabic{figure}}

\begin{figure}[t]
    \includegraphics[width=0.75\linewidth]{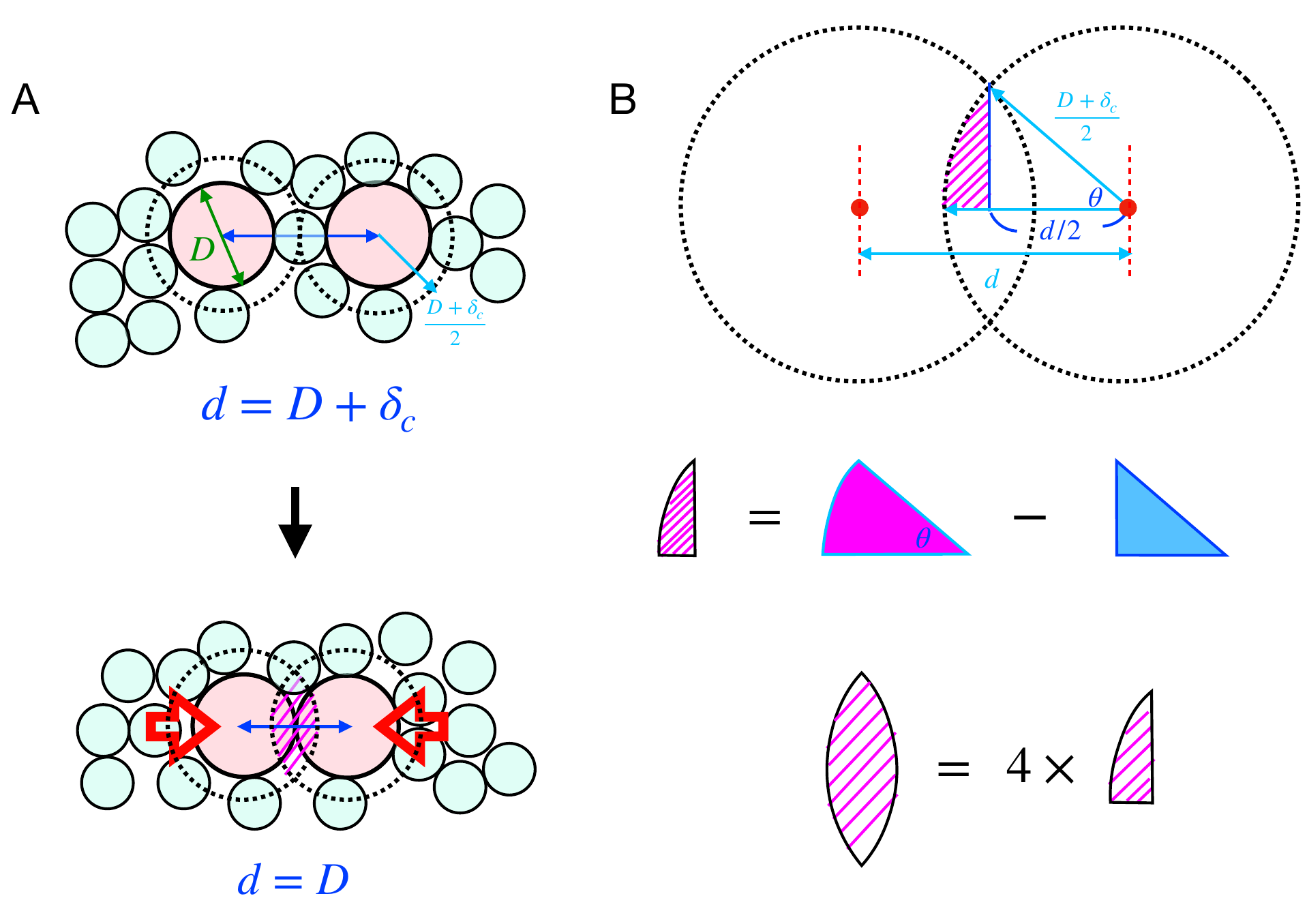}
     \caption{(A) The excess area $A_{\rm ex}(d;\delta_c)$ when $d=D$, marked with hashed lines, created by bringing the two cylinders together. 
     (B) Calculation of the excess area when the inter-cylinder distance is $D<d<(D+\delta_c)$.  
     }
     \label{fig:area}
\end{figure}

\subsection{Derivation of the excess area $A_{\rm ex}(d;\delta_c)$ in Eq.~2}
The excess area created upon bringing the two cylinders at the inter-cylinder distance $D<d<(D+\delta_c)$ (Fig.~\ref{fig:area}A) can be obtained by the geometric consideration shown in Fig.~\ref{fig:area}B.  It is four times the difference in area between the following two shapes: the sector with angle $\theta$ shown in magenta and the right triangle shown in pale blue that fits inside the sector.         
  \begin{itemize}
\item   The area of the sector (colored in magenta in Fig.~\ref{fig:area}B): 
\begin{align}
\pi\left(\frac{D+\delta_c}{2}\right)^2\times \frac{\theta}{2\pi}
\end{align}
with $\theta=\cos^{-1}\left(\frac{d}{D+\delta_c}\right)$. 
\item The area of the triangle (colored in pale blue in Fig.~\ref{fig:area}B): 
\begin{align}
\frac{1}{2}\left(\frac{d}{2}\right)\sqrt{\left(\frac{D+\delta_c}{2}\right)^2-\left(\frac{d}{2}\right)^2}
\end{align}
   \end{itemize}
   Thus, the excess area at $D<d<D+\delta_c$ is 
   \begin{align}
A_{\rm ex}(d;\delta_c)&=4\times \left[\frac{1}{2}\left(\frac{D+\delta_c}{2}\right)^2 \cos^{-1}\left(\frac{d}{D+\delta_c}\right)-\frac{1}{2}\left(\frac{d}{2}\right)\sqrt{\left(\frac{D+\delta_c}{2}\right)^2-\left(\frac{d}{2}\right)^2}\right]\nonumber\\
&=\frac{1}{2}\left[\left(D+\delta_c\right)^2 \cos^{-1}{\left(\frac{d}{D+\delta_c}\right)}-d\sqrt{\left(D+\delta_c\right)^2-d^2}\right]\nonumber\\
&=\frac{D^2}{2}\left[\left(1+\frac{\delta_c}{D}\right)^2\cos^{-1}{\left(\frac{d/D}{1+\delta_c/D}\right)}-\left(\frac{d}{D}\right)^2\sqrt{\left(\frac{1+\delta_c/D}{d/D}\right)^2-1}\right]
   \end{align}
   By rescaling $d$ and $\delta_c$ by $D$, i.e., $x=d/D$ and $\lambda_c=\delta_c/D$, one obtains the expression in Eq.~2 for $1<x<1+\lambda_c$\textcolor{red}{:} 
   \begin{align}
A_{\rm ex}(x;\lambda_c)=\frac{D^2}{2}\left[\left(1+\lambda_c\right)^2 \cos^{-1}{\left(\frac{x}{1+\lambda_c}\right)}-x^2\sqrt{\left(\frac{1+\lambda_c}{x}\right)^2-1}\right]
   \end{align}
   
(i) For $\lambda_c=\delta_c/D\ll 1$, the excess area calculated at $x=1$ (or $d=D$) is approximated as follows. 
 \begin{align}
 A_{\rm ex}(1;\lambda_c)&=\frac{D^2}{2}\left[\left(1+\lambda_c\right)^2 \underbrace{\cos^{-1}{\left(\frac{1}{1+\lambda_c}\right)}}_{\approx \sqrt{2}\lambda_c^{1/2}-\frac{5}{6\sqrt{2}}\lambda_c^{3/2}+\mathcal{O}(\lambda_c^{5/2})}-\underbrace{\sqrt{\left(1+\lambda_c\right)^2-1}}_{\approx \sqrt{2}\lambda_c^{1/2}+\frac{\sqrt{2}}{4}\lambda_c^{3/2}+\mathcal{O}(\lambda_c^{5/2})}\right]\nonumber\\
& \approx \left(\frac{21\sqrt{2}-10}{24\sqrt{2}}\right)D^2\lambda_c^{3/2}+\mathcal{O}(\lambda_c^{5/2})\nonumber\\
&\sim D^{1/2}\delta_c^{3/2}. 
 \end{align}
 
{\color{black}(ii) For $\lambda_c=\delta_c/D\gg 1$, on the other hand, the excess area is approximated as 
\begin{align}
A_{\rm ex}(1;\lambda_c)&=\frac{D^2}{2}\left[\left(1+\lambda_c\right)^2 \underbrace{\cos^{-1}{\left(\frac{1}{1+\lambda_c}\right)}}_{\approx\frac{\pi}{2}-\frac{1}{\lambda_c}+\mathcal{O}(\lambda_c^{-2})}-\underbrace{\sqrt{\left(1+\lambda_c\right)^2-1}}_{\approx (1+\lambda_c)-\frac{1}{2\lambda_c}+\mathcal{O}(\lambda_c^{-2})}\right]\nonumber\\
& \approx (\pi D^2/4)\lambda_c^{2}+\mathcal{O}(\lambda_c)\nonumber\\
&\sim \frac{\pi}{4}\delta_c^2
\end{align}
}

\subsection{Self-Consistent Field Calculation}
To carry out the self-consistent field theory (SCFT) calculation for our mobile brush system with inclusions, we numerically solve the following modified diffusion equations for the chain propagators, $q({\bf r},s)$ and $q^{\dagger}({\bf r},s)$~\cite{deGennesbook,fredrickson2002field,Matsen2006},
\begin{align}
\frac{\partial q({\bf r},s)}{\partial s}&=\left[\frac{R_0^2}{6}\nabla^2-w({\bf r})\right]q({\bf r},s),\nonumber\\
-\frac{\partial q^{\dagger}({\bf r},s)}{\partial s}&=\left[\frac{R_0^2}{6}\nabla^2-w({\bf r})\right]q^{\dagger}({\bf r},s),
\label{eqn:SCFT}
\end{align}
under the one-end grafted boundary condition,  
$q({\bf r},0)=R_0\delta(z-\epsilon)$ with $q^{\dagger}({\bf r},1)=1$. 
The propagator $q({\bf r},s)$ is defined for polymer segments spanning the position along the chain from $0$ to $s$ ($1$ to $s$ for the complementary propagator $q^{\dagger}({\bf r},s)$) for a given effective potential (or self-consistent field) $w({\bf r})$.
The $q({\bf r},s)$ and $q^{\dagger}({\bf r},s)$ are related with the partition function $Q[w]$ and mean monomer density field at ${\bf r}$, $\phi({\bf r})$ as follows: 
\begin{align}
Q[w]=\int d{\bf r}q({\bf r},s)q^{\dagger}({\bf r},s)
\end{align}
and
\begin{align}
\phi({\bf r})=\frac{V_p}{Q[w]}\int_0^{1} dsq({\bf r},s)q^{\dagger}({\bf r},s).
\end{align}
For brush systems with inclusions, Eq.~\ref{eqn:SCFT} is solved by setting $q({\bf r}, s)=0$ and $q^{\dagger}({\bf r}, s)=0$ for ${\bf r}$ inside the inclusions which is equivalent to imposing $w({\bf r})=\infty$ in the interior of the inclusions. 

In SCFT, the self-consistent field $w({\bf r})$ is related to the mean local density field ($\phi({\bf r})$) as 
\begin{align}
w({\bf r})=\Lambda\phi({\bf r}).
\end{align} 
where $\Lambda$ corresponds to the excluded-volume parameter. 
Once $w({\bf r})$ is determined self-consistently, the free energies of brush system with varying separations ($d$) between two inclusions can be calculated using 
\begin{align}
\frac{\beta F(d)}{\sigma R_0^2}=-\frac{A}{R_0^2}\log{\frac{Q[w;d]}{V}}-\frac{A}{2R_0^2V_p}\int d{\bf r}w({\bf r};d)\phi({\bf r};d). 
\label{eqn:Free_energy}
\end{align}
The partition function ($Q[w;d]$) and fields ($w({\bf r};d)$ and $\phi({\bf r};d)$) determined from Eq.\ref{eqn:SCFT} 
vary with the boundary condition defined by 
the shape of inclusions and their distance ($d$). 
The solutions of the SCFT calculation are  plugged into Eq.\ref{eqn:Free_energy} to decide the free energy of the brush system. 

{\color{black}
\subsection*{Implicit vs. explicit brush model}
{\bf Preparation and Simulation of the implicit brush model. }
The potential of mean force (PMF) data obtained from the umbrella sampling were 
processed and converted into a tabulated potential for MD simulations. 
For smooth derivatives for PMF, the data was smoothed using the cubic univariate spline interpolation.  

To provide adequate coverage at short inter-particle distances where umbrella sampling data are sparse due to strong repulsive interactions, the potential was extrapolated inward by $1.0b$ from the minimum distance in the original dataset ($b$ representing the length unit will be omitted hereafter). 
This extrapolation prevents particle overlap and eliminates instabilities during the MD simulations. 

To ensure $C^2$ continuity at both switching boundaries for stable energy conservation in MD simulations, we introduced a switching function $S(r)$, so that it prevents force discontinuities that could destabilize the numerical integration: 
\begin{equation}
S(r) = \begin{cases}
1 & \text{if } r \leq r_{\text{switch}} \\
\dfrac{(r_{\text{cut}}^2 - r^2)^2(r_{\text{cut}}^2 + 2r^2 - 3r_{\text{switch}}^2)}{(r_{\text{cut}}^2 - r_{\text{switch}}^2)^3} & \text{if } r_{\text{switch}} < r < r_{\text{cut}} \\
0 & \text{if } r \geq r_{\text{cut}}
\end{cases}
\end{equation}
%
%
where $r_{\text{switch}} = 22.0$ and $r_{\text{cut}} = 24.0$  were chosen based on the PMF data range and computational efficiency. The behavior of $S(r)$, characterized with the smooth transition from the unity to zero over the switching region and its derivative that ensures force continuity, is shown in Fig.~\ref{fig:switching_function}A.  

The tabulated potential, discretized into 1001 equally-spaced points spanning from the extrapolated minimum distance to the cutoff distance, stores the potential energy $U(r) = U_{\text{PMF}}(r) \cdot S(r)$ and the corresponding force $F(r) = -\partial U(r)/\partial r=-[U'_{\text{PMF}}(r) \cdot S(r) + U_{\text{PMF}}(r) \cdot S'(r)]$. 
Figure~\ref{fig:switching_function}B clarifies the full pipeline of processing the PMF data, showing the original PMF data, the extrapolated region, and the final potential with corresponding forces.

\begin{figure}[ht!]
\centering
\includegraphics[width=0.9\linewidth]{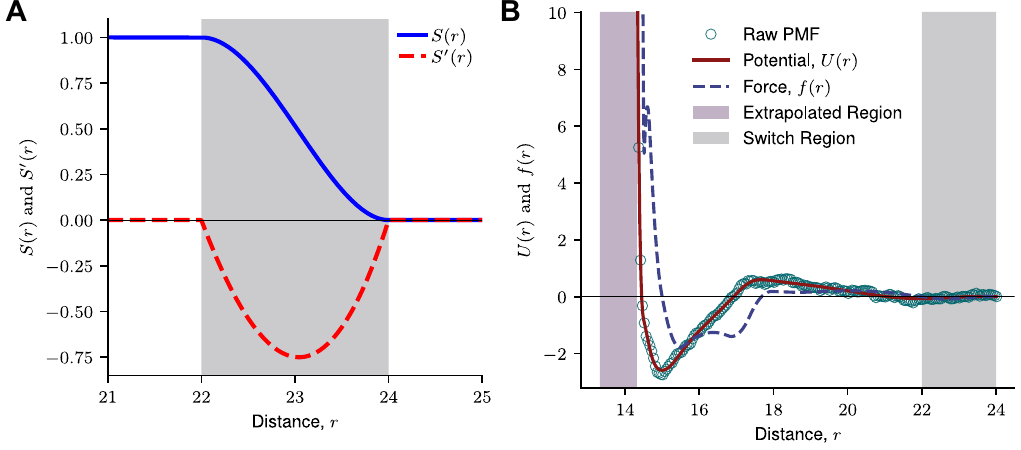}
\caption{(A) The switching function $S(r)$ and its derivative $S'(r)$. The switching function smoothly transitions from 1 to 0 between $r_{\mathrm{switch}}=22.0$ and $r_{\mathrm{cut}}=24.0$, with its derivative ensuring the continuity of force. 
(B) Processed PMF potential and force curves for LAMMPS tabulation. Raw PMF data (green circles) are shown alongside the inward-extrapolated region (light purple shading), the final switched potential $U(r)$ (solid red line), and the corresponding force $f(r)$ (dashed blue line). The gray band denotes the distance range over which the switching function is applied.}
\label{fig:switching_function}
\end{figure}

The MD simulations were performed using LAMMPS with Langevin dynamics to maintain a constant temperature, employing a two-dimensional system of $N_c= 140$ disks with a radius $R_c=7.07$ in an $(L\times L)$-periodic box with $L= 331.6$, which gives rise to the area density of $\phi_c(=N_c\times \pi R_c^2/L^2)\approx 0.2$.  
The system was simulated for a total time of $6 \times 10^6 \tau$ using a time step of $\delta t = 0.01\tau$. 
The production sampling was conducted using only the latter half of the simulation.
\\

{\bf Preparation and Simulation of the explicit brush model. }
For comparison with the implicit brush model, 
the same number ($N_c=140$) of composite rigid-body cylinders was inserted into an identical periodic simulation box. 
To enhance the configurational sampling of the entire system, we rescaled the total cylinder mass to the unity, and we simulated 10 independent runs with different initial configurations. 
After an initial relaxation period of $10^5\tau$, the remaining trajectories ($\sim O(10^5)\tau$) were used for data collection.
\\

{\bf Clustering. }
In both implicit and explicit brush models, we consider that disks or cylinders whose center-of-mass separation ($r_{ij}$) satisfies $r_{ij}<D+2.5b$ are in the same cluster. 
}


\bibliography{mybib1,jwyu}

\end{document}